\documentclass[twocolumn,prl,showpacs,superscriptaddress]{revtex4}
\usepackage{graphicx}
\usepackage{amsmath}
\usepackage{amssymb}

\begin{document}
\title{Scanning Tunneling Microscopy studies on CeCoIn$_5$ and
CeIrIn$_5$}

\author{S.~Ernst}
\affiliation{Max Planck Institute for Chemical Physics of Solids,
N\"othnitzer Stra\ss e~40, 01187 Dresden, Germany}
\author{S.~Wirth}
\affiliation{Max Planck Institute for Chemical Physics of Solids,
N\"othnitzer Stra\ss e~40, 01187 Dresden, Germany}
\author{F.~Steglich}
\affiliation{Max Planck Institute for Chemical Physics of Solids,
N\"othnitzer Stra\ss e~40, 01187 Dresden, Germany}
\author{Z.~Fisk}
\affiliation {Department of Physics and Astronomy, University of
California, Irvine, CA 92697, USA}
\author{J.~L.~Sarrao} \affiliation {Los Alamos National
Laboratory, Los Alamos, NM 87545, USA}
\author{J.~D.~Thompson} \affiliation {Los Alamos National
Laboratory, Los Alamos, NM 87545, USA}
\date{\today}

\begin{abstract}
High--quality single crystals of the heavy fermion superconductors
CeCoIn$_5$ and CeIrIn$_5$ have been studied by means of
low--temperature Scanning Tunneling Microscopy. Methods were
established to facilitate \textit{in-situ} sample cleaving.
Spectroscopy in CeCoIn$_5$ reveals a gap which persists to above
$T_c$, possibly evidencing a precursor state to SC. Atomically
resolved topographs show a rearrangement of the atoms at the
crystal surface. This modification at the surface might influence
the surface properties as detected by tunneling spectroscopy.
\end{abstract}
 \maketitle

\section{Introduction}During the past decade, the Ce\textit{M}In$_\text{5}$
(\textit{M} = Co, Ir, Rh) family of heavy fermion (HF) compounds
has attracted considerable scientific interest. The latter was
largely motivated by the discovery of unconventional
superconductivity (SC) in these materials
\cite{hegger:00,petrovic:01a,petrovic:01b,sarrao:07}. Unlike in
conventional superconductors, where tiny amounts of magnetic
impurities suppress  the superconducting state, HF
superconductivity develops in an inherently magnetic environment.
For instance, it was shown for the case of the prototypical HF
superconductor CeCu$_\text{2}$Si$_\text{2}$ that the complete
occupancy of the corresponding lattice sites by the magnetic
Ce$^{3+}$ ions is necessary to generate SC \cite{steglich:79}. It
is generally accepted that magnetic fluctuations may potentially
induce electron pairing into a superconducting condensate
\cite{schmitt-rink:86,scalapino:86,marthur:98}. The precise
interplay between magnetism and SC in HF compounds, however, is
not yet understood in all detail. The knowledge of the excitation
spectrum of the superconducting quasiparticles is one crucial
aspect for the understanding of the superconducting state.

In this paper we report the investigation of the HF
superconductors CeCoIn$_\text{5}$ and CeIrIn$_\text{5}$ by means
of Scanning Tunneling Microscopy (STM). An important and powerful
feature of STM is the possibility to perform local spectroscopy on
the atomic scale, referred to as Scanning Tunneling Spectroscopy
(STS). By measuring the tunneling current at constant tip-sample
distance as a function of the applied voltage, one can probe
directly the electronic density of states (DOS) of the sample, and
thereby obtain information on the charge degrees of freedom of the
material investigated. The accessibility of SC at ambient
pressure, as well as the -- for HF compounds -- comparably high
transition temperatures (2.3\,K and 0.4\,K, respectively) render
CeCoIn$_\text{5}$ and CeIrIn$_\text{5}$ ideal candidates to study
HF superconductivity by STM. In both materials, the
superconducting state is unconventional. The gap function has line
nodes, and an order parameter with $d$-wave symmetry has been
suggested \cite{izawa:01,vorontsov:06}. Although the ground states
are non-magnetic, strong antiferromagnetic fluctuations are
present\,\cite{kohori:01}, and it has been speculated that these
fluctuations mediate the superconducting pairing
\cite{stock:08,kasahara:09}. Together with other experimental
findings, the unconventional nature of SC and its close
relationship to magnetism has lead to the suggestion that the
1:1:5 compounds are remarkably similar to the cuprate High-$T_c$
superconductors \cite{nagajima:07}.

We present STM/STS data obtained on single crystalline 1:1:5
samples. A gap compatible with $d$-wave SC is observed in the
conductance spectra of CeCoIn$_\text{5}$. The presence of a
gap-like feature in a temperature range above $T_c$ might evidence
a precursor state similar to the pseudogap phase in the underdoped
cuprates. Based on atomically resolved topography data, the
possible influence of a modified surface structure on STS is
discussed.

\section{Experimental Details}
The application of STM and, in particular,
STS to HF superconductors has only become feasible very recently.
One reason is that the involved energy scales are in the order of,
or even less than 1\,meV. This calls for both an excellent energy
resolution and the accessibility of temperatures considerably
lower than the superconducting transition temperatures $T_c$. Our
tunneling experiments were conducted in a $^3$He cryostat with a
base temperature of 320\,mK at the STM. In order to maintain an
unperturbed vacuum tunnel junction over a sufficiently long period
of time, the STM operates under UHV conditions
($p$\,$<$\,$10^{-10}$\,mbar)\,\cite{omicron}. An adequate energy
resolution ($\approx$\,80\,$\mu$V) of the STM has been verified by
investigating the superconducting gap of Al. We used
electrochemically etched tungsten tips which were conditioned
\textit{in situ} by Ne ion sputtering and electron beam annealing
\cite{ernst:07}.

In order to obtain a clean tunnel junction in the first place,
several methods for surface preparation have been applied to the
single crystals. Samples treated by etching in HCl, by \textit{in
situ} radiative annealing, by Ar ion sputtering, or by a
combination of these methods, exhibited extensive structural
damage at the surface, including indium segregation. Therefore,
such samples did not yield any reasonable STM results. However,
both materials could successfully be cleaved \textit{in situ} at
room temperature. CeCoIn$_\text5$ forms platelet--like crystals.
The samples used in our experiments had a size of few hundred
$\mu$m along the \textit{c}--axis, and up to 2\,mm within the
\textit{ab}--plane. In order to facilitate the cleaving, a
stainless steel post was glued onto a sample using strong
epoxy\,\cite{epoxy}. The post was then torn off \textit{in situ},
often ripping off part of the sample and exposing a fresh surface.
Since CeCoIn$_\text5$ is rather ductile, it was found to be
important to pull off the post straight in order to avoid shear
forces while cleaving. The highest success rate was achieved when
the post was polished at the face used for glueing. Furthermore,
the shape of the post should be well adapted to the sample
dimensions. In contrast, CeIrIn$_\text5$ crystals are rather
brittle and larger along the \textit{c}--axis. These samples could
be cleaved directly, using an \textit{in situ} cleaving tool. The
latter consists of two tungsten--carbide jaws that form a pair of
pliers. Considering the layered crystal structure of the 1:1:5
materials, one might expect that the samples can easily be cleaved
parallel to the \textit{ab}--plane. However, the samples often did
not cleave, but broke in an irregular fashion. In spite of the
macroscopic roughness of the resulting surfaces it was still
possible to perform tunneling experiments on microscopically small
areas.

\section{Results}
\begin{figure}[t]
\centering
\includegraphics[width=8.4cm]{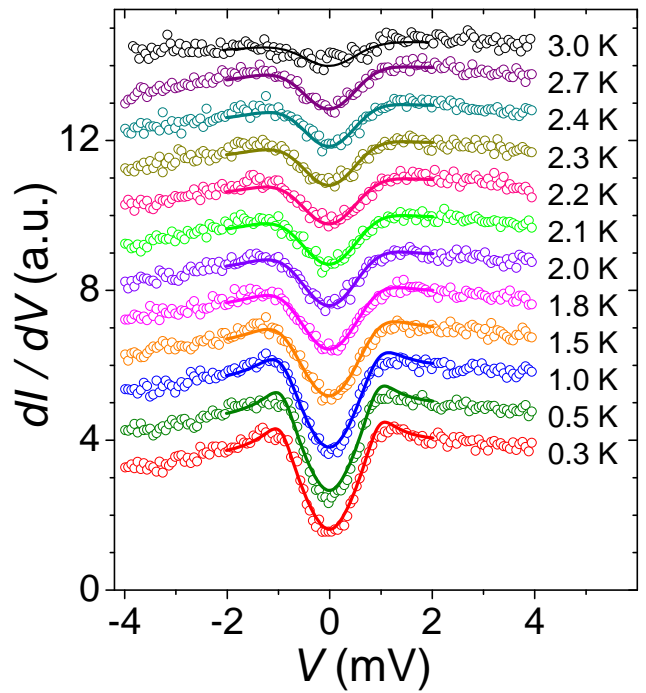} \caption{
Differential conductance data obtained on CeCoIn$_\text{5}$ (open
circles) and fits according to BCS theory (lines, see text). For
clarity, the curves are shifted vertically.} \label{sts}
\end{figure}

\begin{figure}[t]
\centering
\includegraphics[width=8.0cm]{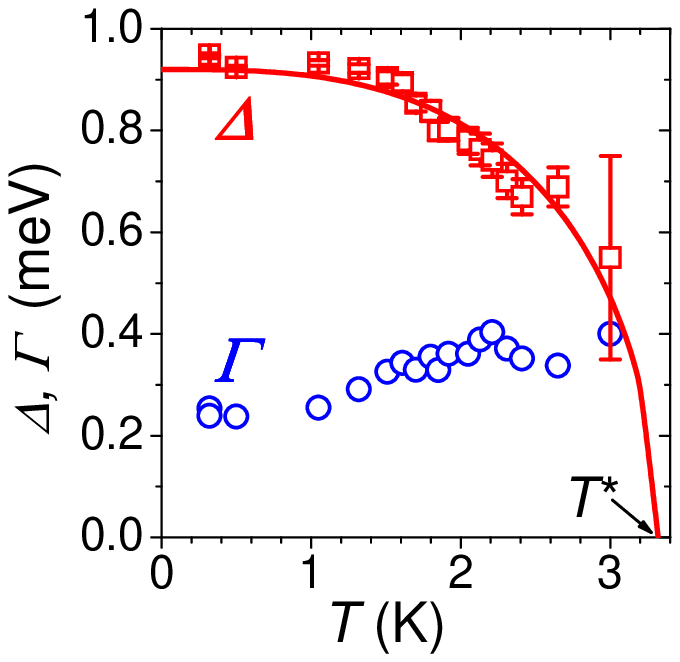} \caption{
The extracted fit parameters $\Delta$, $\Gamma$ vs. temperature.
The solid line is a fit for BCS theory for nodal SC. $T^*$ marks
the temperature at which $\Delta(T)=0$ according to the fit.}
\label{gamma}
\end{figure}

The primary objective of this work was to obtain spectroscopic
information in the superconducting state of the 1:1:5 compounds. A
gap was observed in the differential conductance spectra of
CeCoIn$_\text{5}$. Figure \ref{sts} shows its temperature
evolution between 320\,mK and 3\,K. The curves were acquired with
a bias voltage of $V_g$\,= 14 mV at a set-point current of
$I_\text{set}=$\,340 pA. $dI/dV$ was measured directly, using a
lock--in amplifier with a modulation amplitude of 70\,$\mu$V at
180\,Hz. For a better visibility the individual curves were
shifted vertically. Upon increasing temperature the zero bias
conductance increases, indicating a closing of the gap. The solid
lines in the figure represent fits to the tunneling conductance,
considering thermal broadening. For the superconducting order
parameter, we assumed $d_{x^2-y^2}$ symmetry, which is likely to
be present in CeCoIn$_\text{5}$ \cite{izawa:01,vorontsov:06}. In
this case, the BCS expression for the superconducting excitation
spectrum takes the form \cite{won:94}
\begin{equation}
\rho(E)\propto\mbox{Re}\int_0^{2\pi}\frac{d\phi}{2\pi}
\frac{E-i\Gamma}{\sqrt{(E-i\Gamma)^2-\Delta^2\cos^2(2\phi)}}\; ,
\end{equation}
where $\Delta$ is the maximum value of the angular dependent gap
function. The additional lifetime broadening parameter $\Gamma$
accounts for in-gap states due to inelastic scattering
\cite{dynes:78}. A linear term has been included to the fit
function in order to approximate the slightly asymmetric
background. The fit parameters $\Delta$ and $\Gamma$ are plotted
{\it vs.} temperature in Fig.\,\ref{gamma}. The error bars are
estimated from the fitting procedure. Expectedly, the
superconducting order parameter $\Delta$ decreases with
temperature, whereas $\Gamma$ increases. The solid line represents
a fit with an approximate expression for the temperature
dependence of the order parameter in nodal superconductors
\cite{dora:01},
\begin{equation}
\Delta(T)=\Delta_0\sqrt{1-(T/T^*)^3}\;\; .
\end{equation}
Here, $T^*$ denotes the temperature at which $\Delta(T)$
extrapolates to zero.

The fit yields a zero-temperature value of the order parameter
$\Delta_0=0.93\,$meV. Similar values were previously found in
point contact spectroscopy experiments \cite{goll:03,rourke:05},
but also different values have been reported \cite{park:05}. The
resulting ratio $2\Delta_0/k_BT_c\approx10$ indicates that
CeCoIn$_\text{5}$ is a strong coupling superconductor. This is
consistent with the large jump detected in the specific heat
\cite{petrovic:01b}. Notably, the gap does not close at
$T_c=2.3$\,K, but persists up to a temperature $T^*=3.3$\,K. This
behaviour is reminiscent of the pseudogap phase of underdoped
cuprates \cite{renner:97}. Considering the striking similarities
to the cuprates, it may be reasonable to expect similar effects in
the 1:1:5's. Indeed, a pseudogap above $T_c$ has been observed in
electrical resistivity measurements on
CeCoIn$_\text{5}$\,\cite{sidorov:02}. The value $T^*=3.3$\,K
extrapolated from our data is in agreement with the pseudogap
temperature $T_{pg}$ reported therein. Also for the other members
of the 1:1:5 family, the existence of a precursor state to SC has
been reported \cite{nair:08,kawasaki:02}. At present, the nature
of the precursor state is still speculative. In the case of
CeIrIn$_\text{5}$, however, there is strong evidence that both SC
and this state have a common origin. The data shown here support
this idea, as the gap evolves continuously at around $T_c$. As
noted in Ref. \,\cite{izawa:01}, the fourfold anisotropy in the
thermal conductivity persists to 3.2\,K. Hence, a possible
precursor state above $T_c$ might have the same $d$-wave symmetry
as the superconducting state.

\begin{figure}[t]
\centering
\includegraphics*[width=8.4cm]{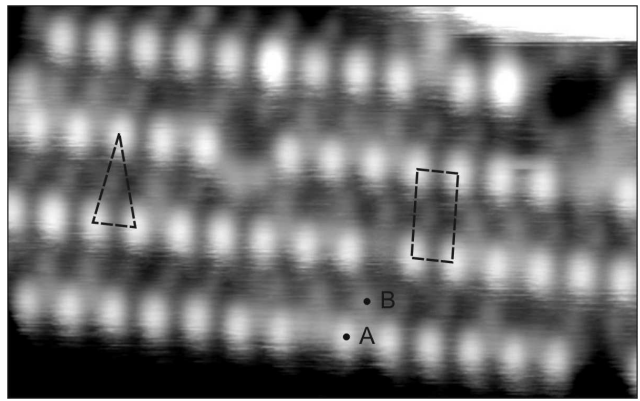}
\caption{STM topography obtained on CeIrIn$_\text5$. The image
covers an area of 6.4$\times$10\,nm$^2$ and a $z$-range of
0.63\,nm, the parameters were $V_g$\,=\,600\,mV,
$I_\text{set}$\,=\,0.3\,nA, $T$\,=\,330\,mK. A and B denote two
kinds of atomic corrugations. The markers illustrate different
arrangements of the atoms of type A.} \label{atoms}
\end{figure}

Throughout our measurements, the gap in the tunneling conductance
had a low spatial reproducibility, and many sample spots did not
show any gap. Also in these cases, a clean vacuum tunnel junction
was verified from the dependence of the tunneling current $I$ on
the tip-sample distance $z$ yielding an effective work function
$\Phi=2.2$\,eV. Hence, the variations over the surface area might
have a different origin, such as changes in the composition or
structure. Even on \textit{in-situ} cleaved samples, the surface
properties as detected by STM might locally differ significantly
from those of the bulk\,\cite{fischer:07}. Topographic data
obtained on CeIrIn$_\text5$ hints toward this possibility.
Figure~\ref{atoms} exemplifies an atomically-resolved topograph
acquired on an \textit{in-situ} cleaved sample. The image reveals
different atomic sites: In addition to the most prominent
corrugations (marked as ``A'' in the figure), other sites can be
identified in between (marked as ``B''). Different arrangements of
the surface atoms are clearly visible within the image. The
corrugations of type A form either a triangular lattice, or a
rectangular one, as indicated by the respective markers. The two
different arrangements merge through dislocations due to a defect.
In order to identify the crystallographic plane visible in the
figure, we compared the data with the lattice dimensions of
CeIrIn$_\text5$. The sample had been mounted parallel to the
\textit{ab}-plane. A high tilting angle of the imaged sample area
of 37 degree with respect to the scanning plane points towards a
plane of low symmetry. This slope, together with the interatomic
distances of the triangular lattice of the A--type atoms,
$(6.7\pm0.3)$\,{\AA}\,and $(14.6\pm0.5)$\,{\AA}, is in reasonable
agreement with the arrangement of the Ce or Ir atoms within the
\{113\} plane of CeIrIn$_\text5$. The corrugations marked as B
could then originate from the Ir or Ce atoms of the subjacent
lattice plane. It should be noted, however, that this is only
one possible interpretation of Fig.~\ref{atoms}. The explanation
is complicated by the fact that the surface atoms are obviously
rearranged, at least part of them. Hence, it is not clear whether
the surface periodicity is at all comparable to the bulk crystal
structure. These surface modifications may occur naturally, or
they might be induced by the applied cleaving procedure
\cite{fischer:07}.

Atomically resolved images were obtained on areas up to
$60\times60\,\text{nm}^2$, revealing narrow terraces of various
lattice planes. Larger flat terraces, reflecting the unperturbed
crystal structure, were not observed. A surface that differs
substantially from the bulk structure may have completely
different properties. In particular, SC may be suppressed close to
the crystal surface. A modified surface structure might,
therefore, hinder the reproducible observation of SC--derived
changes to the density of states by STS. Similar arguments might
apply to the case of CeCoIn$_\text{5}$ since it has the same
tetragonal crystal structure as its Ir counterpart. Possibly, one
might even encounter similar difficulties in other HF
superconductors. Therefore, dedicated surface preparation methods
may have to be developed for each material in order to overcome
such  cleaving--related problems.

\section{Conclusions}
In conclusion, we reported on systematic investigation of
\textit{in situ} cleaved Ce\textit MIn$_\text5$, \textit M = Co,
Ir, single crystals by means of low--temperature STM/STS. A gap
detected in CeCoIn$_\text{5}$ is compatible with $d_{x^2-y^2}$
symmetry of the superconducting order parameter. The persistence
of the gap beyond $T_c$ might evidence a precursor state to SC,
similar to the pseudogap observed in underdoped cuprates.
Atomically resolved topography imaging reveals a rearrangement of
the surface atoms in CeIrIn$_\text{5}$. This modification might
influence the tunneling spectroscopy not only in this compound,
but also in CeCoIn$_\text{5}$ or even in other HF superconductors.

\section{Acknowledgement}
Work in Dresden was supported by the DFG research unit 960
"Quantum phase transitions". Z.~F. acknowledges support through
NSF-DMR-071042.

\end{document}